\newcounter{myctr}
\def\myitem{\refstepcounter{myctr}\bibfont\noindent\ifnum\themyctr>9\else\phantom{0}\fi\hangindent17pt\themyctr.\enskip}

\documentclass{ws-ijqi}

\begin{document}

\markboth{D. Ciampini, O. Morsch, E. Arimondo}
{QUANTUM CONTROL IN STRONGLY DRIVEN OPTICAL LATTICES}

%%%%%%%%%%%%%%%%%%%%% Publisher's Area please ignore %%%%%%%%%%%%%%
\catchline{}{}{}{}{}
%%%%%%%%%%%%%%%%%%%%%%%%%%%%%%%%%%%%%%%%%%%%%%%%%%%%%%%%%%%%%%%%%%%

\title{QUANTUM CONTROL IN STRONGLY DRIVEN OPTICAL LATTICES}

\author{DONATELLA CIAMPINI}

\address{CNISM and INO-CNR, Dipartimento di Fisica,
Universit\`{a} di Pisa, L.go B. Pontecorvo 3\\
Pisa, I-56127, Italy\\
ciampini@df.unipi.it}

\author{OLIVER MORSCH}

\address{INO-CNR, Dipartimento di Fisica,
Universit\`{a} di Pisa, L.go B. Pontecorvo 3\\
Pisa, I-56127, Italy\\
morsch@df.unipi.it}

\author{ENNIO ARIMONDO}

\address{CNISM and INO-CNR, Dipartimento di Fisica,
Universit\`{a} di Pisa, L.go B. Pontecorvo 3\\
Pisa, I-56127, Italy\\
arimondo@df.unipi.it}

\maketitle

\begin{history}
\received{Day Month Year}
\revised{Day Month Year}
%\accepted{Day Month Year}
%\comby{(xxxxxxxxxx)}
\end{history}

\begin{abstract}

Matter waves can be coherently and adiabatically loaded and controlled in strongly driven optical lattices. This coherent control is used in order to modify the modulus and the sign of the tunneling matrix element in the tunneling Hamiltonian. Our findings pave the way for studies of
driven quantum systems and new methods for engineering Hamiltonians that are impossible to realize with static techniques.

\end{abstract}


\section{Introduction}	
The experimental realization of quantum degenerate states in ultracold
atoms gases in the 1990's opened the possibility to coherently control many-body systems
isolated from external perturbations and at temperatures close to absolute
zero. These techniques, applied to ultracold atomic gases in optical lattices~\cite{morsch_review}, allow physicists to operate at the interface between quantum information
and many-body systems~\cite{Jaksch_1999,Lew_rev}.

The possibility of easily changing the parameters (depth, lattice
constant etc.) of a defect-free periodic potential permits one to observe, for example,
Bloch oscillations~\cite{noi2001} and the quantum phase transition into a Mott-insulator
regime~\cite{fisher_1989,jaksch_98,greiner}.
In strongly driven quantum systems, a time-dependent (in particular periodic) perturbation is introduced into the system in order to probe it or to change its fundamental properties. From a theoretical point of view, under certain circumstances
the phenomena resulting from the strong driving can be described by absorbing the effect of the external driving inside one of the
parameters of the unperturbed system. This approach bears a strong resemblance to the renormalization of the mass of an electron in a the periodic potential of a crystal familiar from condensed matter physics.

Here we present experimental results on Bose-Einstein condensates in driven optical lattices. We show that by periodically forcing the system, the fundamental tunneling properties of the atoms inside the lattice can be adiabatically changed and the system can be carried
into novel regimes impossible to reach with static techniques.

In the following Sec.~2 we review the theoretical description of strongly driven optical lattices.
In Sec.~3 we describe the experimental setup and the details of how the driving is implemented and report our measurements, which demonstrate the renormalization of the tunneling matrix element and the dynamics of the condensate after an abrupt change of the sign of the tunneling.  In the final Sec.~4 we spell out our conclusions.

\section{Theory}

Our system consisting of a Bose-Einstein condensate inside a
sinusoidally shaken one-dimensional optical lattice is
approximately described by the Hamiltonian
\begin{equation}
\hat{H}_0=-J\sum_{\langle
i,j\rangle}(\hat{c}_i^\dagger\hat{c}_j+\hat{c}_j^\dagger\hat{c}_i)+\frac{U}{2}\sum_j
\hat{n}_j(\hat{n}_j-1)+K\cos(\omega t)\sum_j j\hat{n}_j,
\label{HamForce}
\end{equation}
where $\hat{c}_i^{(\dagger)}$ are the boson creation and
annihilation operators on site $i$,
$\hat{n}_i=\hat{c}_i^{\dagger}\hat{c}_i$ are the number operators,
and $K$ and $\omega$ are the strength and angular frequency of the
shaking, respectively. The first two terms in the Hamiltonian
describe the Bose-Hubbard model~\cite{jaksch_98} with the
tunneling matrix element $J$ and the on-site interaction term $U$.
By using the Houston functions~\cite{Holthaus:1999:book} (a different procedure leading to the same result was used by Ref.~\cite{dunlap:1986:prb}) it can be demonstrated that the presence of a driving force corresponds to a renormalization of the Bose-Hubbard Hamiltonian, leading to an effective tunneling parameter~\cite{eckardt_05,creffield_06}
\begin{equation}
J_\mathrm{eff}=J\mathcal{J}_0(K_0),
\label{bessel}
\end{equation}
where $\mathcal{J}_0$ is the zeroth-order ordinary Bessel function
and we have introduced the dimensionless parameter
$K_0=K/\hbar\omega$.

\section{Experiment}
\subsection{Setup}
\begin{figure}[pb]
\centerline{\psfig{file=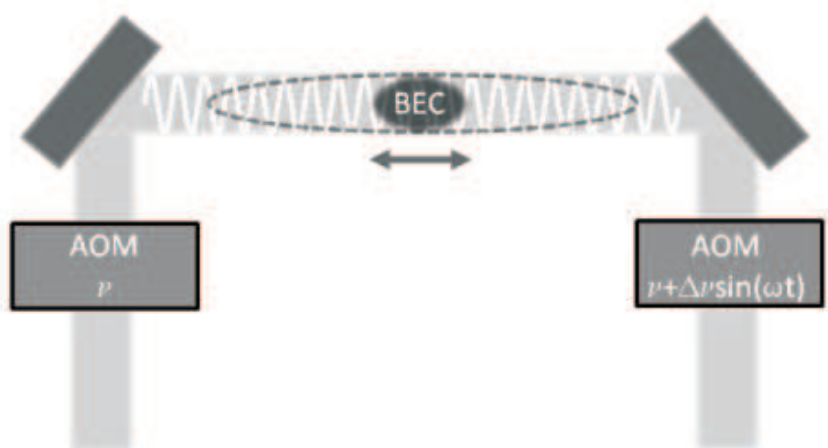, width=12.0cm}}
\caption{Experimental setup for a driven optical lattice. The two beams
originating from the same laser pass trough two separate acousto-optic
modulators (AOM). Each AOM is driven by a function generator whose frequency can be modulated both with internal
preset functions or by using a triggered external channel.}
\label{Setup}
\end{figure}

In our experiment we created BECs of about $5\times 10^4$
87-rubidium atoms in an optical crossed dipole trap (for details of the experiment see~\cite{lignier_2007,zenesini_2009}). After obtaining pure condensates, the powers of the trap beams were adjusted in order to
obtain elongated condensates with the desired trap frequencies
($\approx 20\,\mathrm{Hz}$ in the longitudinal direction and
$80\,\mathrm{Hz}$ radially). Along the axis of one of the dipole
trap beams a one-dimensional optical lattice potential was then
added by ramping up the power of the lattice beams in
$50\,\mathrm{ms}$ (the ramping time being chosen such as to avoid
excitations of the BEC). The optical lattices used in our
experiments were created using two counter-propagating gaussian
laser beams ($\lambda = 852\,\mathrm{nm}$) with $120\,\mathrm{\mu
m}$ waist and a resulting optical lattice spacing $d_L= \lambda /2
= 0.426 \,\mathrm{\mu m}$, sketched in Fig.~\ref{Setup}.

The depth $V_0$ of the resulting
periodic potential is measured in units of $E_{\rm rec}= \hbar^2
\pi^2 / (2m d_L^2)= p_{\rm rec}^2 /2m$, where $m$ is the mass of the Rb atoms and $p_{\rm rec}$ is the recoil momentum. By
introducing a frequency difference $\Delta \nu$ between the two
lattice beams (using acousto-optic modulators which also control
the power of the beams), the optical lattice could be moved at a
velocity $v=d_L\Delta \nu$ or accelerated with an acceleration
$a=d_L\frac{d\Delta\nu}{dt}$. In order to periodically shake the
lattice, $\Delta \nu$ was sinusoidally modulated with angular
frequency $\omega$, leading to an inertial
time-varying force in the lattice reference frame
\begin{equation}
F(t)= m\omega d_L \Delta\nu_{\mathrm{max}}\cos(\omega t)=
F_{\mathrm{max}}\cos(\omega t).
\end{equation}
The peak shaking force $F_{\mathrm{max}}$ is related to the
shaking strength $K$ appearing in Eq.~\ref{bessel} by $K=F_{\mathrm{max}}d_L$, and hence the dimensionless shaking parameter is
\begin{equation}
K_0=\frac{K}{\hbar \omega} =
\frac{md_L^2\Delta\nu_{\mathrm{max}}}{\hbar}=\frac{\pi^2\Delta\nu_{\mathrm{max}}}{2\omega_{\mathrm{rec}}}.
\end{equation}
For a typical shaking frequency $\omega/2\pi = 3\,\mathrm{kHz}$
the spatial shaking amplitude $\Delta x_{\mathrm{max}}\approx 0.5 d_L$ at $K_0=2.4$.

\subsection{Results}

\begin{figure}[pb]
\centerline{\psfig{file=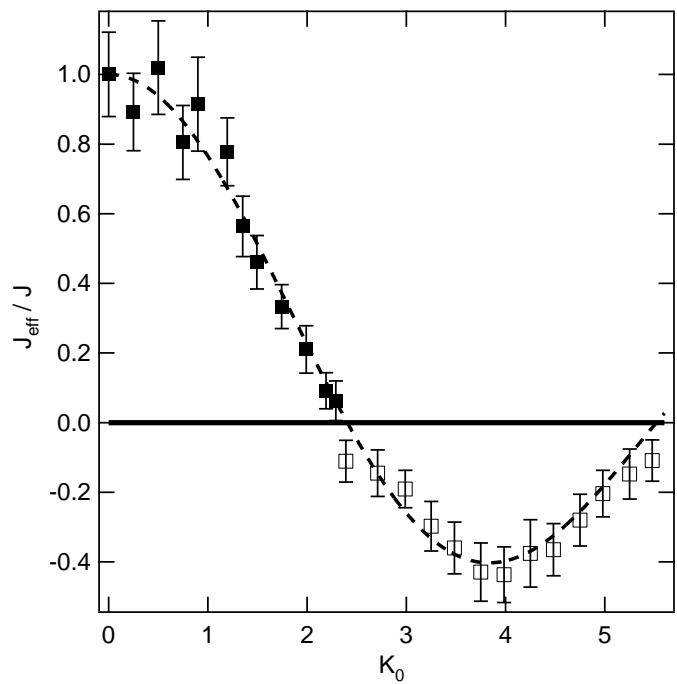, width=10.0cm}}
\vspace*{8pt}
\caption{Dynamical suppression of tunneling in an
optical lattice. Shown here are the values of
$J_{\mathrm{eff}}/J$ calculated from the expansion velocities
as a function of the shaking parameter $K_0$. The sign of $J_{\mathrm{eff}}/J$ has been determined by time-of-flight experiments. The lattice depth
and shaking frequency were: $V_0/E_{\mathrm{rec}}=6$,
$\omega/2\pi = 1\,\mathrm{kHz}$. The dashed line is the theoretical
prediction.}
\label{Bessel1}
\end{figure}
After loading the BECs into the optical lattice, the frequency
modulation of one of the lattice beams creating the shaking was
suddenly switched on. In order to measure the
effective tunneling rate $|J_{\mathrm{eff}}|$ between the lattice
wells (where the modulus indicates that we are not sensitive to
the sign of $J$, in contrast to the time-of-flight experiments
described below), we then switched off the dipole trap beam that
confined the BEC along the direction of the optical lattice,
leaving only the radially confining beam switched on (the trap
frequency of that beam along the lattice direction was on the
order of a few Hz and hence negligible on the timescales of our
expansion experiments, which were typically less than
$200\,\mathrm{ms}$). The BEC was now free to expand along the
lattice direction through inter-well tunneling and its {\it
in-situ} width was measured using a resonant flash, the shadow
cast by which was imaged onto a CCD chip.

\begin{figure}[pb]
\centerline{\psfig{file=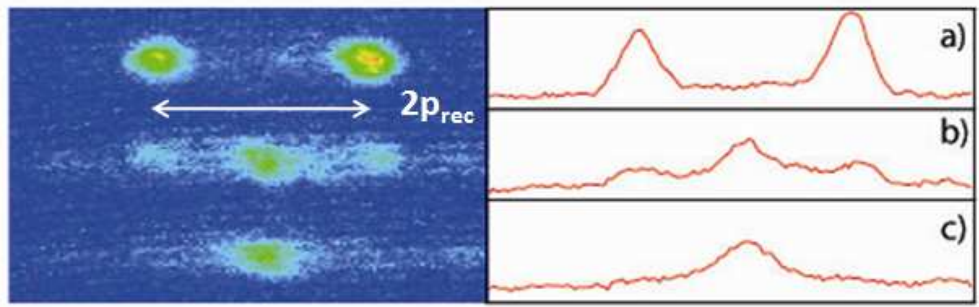, width=14.5cm}}
\vspace*{8pt}
\caption{Interference pattern of a time-of-flight experiment with final
acceleration to the zone edge is shown for various times in the modulated lattice at $K_0 = 4$ (where $J_{\mathrm{eff}}$is negative): (a) t= 1 ms, (b) t = 2 ms, (c) t = 2.5 ms (on the $x$-axis, the
spatial position has been converted into the corresponding
momentum in units of the recoil momentum $p_\mathrm{rec}$.}
\label{NegJ}
\end{figure}

The sign of $J_{\mathrm{eff}}$ can be measured by looking at the interference pattern of a BEC released from a modulated lattice, after a sufficiently long time of flight.
After shaking the condensate in the lattice for a fixed time between $1$
and $\approx 200\,\mathrm{ms}$ we accelerated the lattice for
$\approx 1\,\mathrm{ms}$ with an acceleration chosen such that at the end of the process
the BEC was in a staggered state at the edge of the Brillouin
zone. After switching off the dipole trap and lattice beams and
letting the BEC fall under gravity for $20\,\mathrm{ms}$, this
resulted in an interference pattern featuring two peaks of roughly
equal height for shaking amplitudes $K_0<2.4$ for which $J_{\mathrm{eff}}>0$. The visibility of the interference pattern reflects the phase coherence of the BEC. In the region between the first
two zeroes of the Bessel function, where $\mathcal{J}_0<0$,
the interference pattern is shifted by half a Brillouin zone, as predicted from
the dependence of the band shape on the sign of $J_{\mathrm{eff}}$. When the shaking is
switched on with $K_0$ between 2.4 and 5.9, the atoms, in fact, occupy the
new ground state at the edge of the Brillouin zone. Fig.~\ref{NegJ} shows how the system evolves from the initial state for which $J_{\mathrm{eff}}>0$ to the new one with $J_{\mathrm{eff}}<0$.

\section{Conclusions}
In summary, we have measured the dynamical suppression of
tunneling of a BEC in strongly shaken optical lattices and found
excellent agreement with theoretical predictions.

The fact that time-of-flight images with high-contrast interference patterns could be realized shows
that the tunneling suppression occurs in a phase-coherent way and
can, therefore, be used as a tool to control the tunnelling matrix
element and without disturbing the condensate. The condensate then occupies a single Floquet state, just as it occupies a single energy eigenstate when there is no forcing. The interference patterns also allowed us to verify that strong driving can lead to a negative value of the tunnelling parameter (impossible to create with static potentials). Such systems with a negative value and/or anisotropy of the tunneling parameter may produce new quantum phases. In a triangular lattice, such a negative tunneling energy can, for instance, mimic frustrated spin systems and also simulate other complicated Hamiltonians that are difficult to study in solid state materials~\cite{Triangular}.

Our results can also be interpreted as a realization of \textit{dressed matter-waves}, where the shaking becomes a \textit{dressing} that can take the system into a  regime where the time-periodical perturbation behaves like a time-independent parameter~\cite{Holthaus_2008}. This feature allows us to simulate regimes impossible to reach by changing the lattice depth, the atomic species etc. At the same time the realization of such a system allows us to study the Floquet states for large numbers of atoms and lattice sites, which is extremely complicated from the computational point of view.

\section*{Acknowledgments}

The authors would like to thank the PhD students and post-docs participating in the experiments described in this article: H. Lignier, C. Sias, A. Zenesini, Y. Singh and J. Radogostowicz. We acknowledge funding by the MIUR (PRIN-2007),  CNISM Progetto d'Innesco 2007, OLAQUI (EC FP6-511057), and NAMEQUAM (EC FP7-225187).

\end{document}